\documentclass[12pt,preprint]{aastex}

\shorttitle{Searching for Variability in M4}
\shortauthors{Ferdman et al.}

\begin{document}

\title{Searching for Variability in the Globular Cluster Messier 4\altaffilmark{1}}


\author{Robert D. Ferdman\altaffilmark{2}, Harvey B. Richer\altaffilmark{2}, James Brewer\altaffilmark{2}, Greg G. Fahlman\altaffilmark{3}, Brad K. Gibson\altaffilmark{4}, Brad M. S. Hansen\altaffilmark{5}, Mark E. Huber\altaffilmark{2}, Rodrigo A. Ibata\altaffilmark{6}, Jasonjot S. Kalirai\altaffilmark{2}, Jaymie M. Matthews\altaffilmark{2}, R. Michael Rich\altaffilmark{5}, Jason F. Rowe\altaffilmark{2}, Michael M. Shara\altaffilmark{7}, and Peter B. Stetson\altaffilmark{3}}

\altaffiltext{1}{Based on observations with the NASA/ESA \textsl{Hubble Space Telescope}, obtained at the Space Telescope Science Institute, which is operated by the Association of Universities for Research in Astronomy, Inc., under NASA contract NAS5-26555.  These observations are associated with proposal GO-8679.}
\altaffiltext{2}{Department of Physics and Astronomy, University of British Columbia, 6224 Agricultural Road, Vancouver, BC V6T 1Z4, Canada; ferdman@astro.ubc.ca, jbrewer@astro.ubc.ca, jkalirai@astro.ubc.ca, richer@astro.ubc.ca, rowe@astro.ubc.ca.}
\altaffiltext{3}{National Research Council, Herzberg Institute of Astrophysics, 5071 West Saanich Road, RR5, Victoria, BC V9E 2E7, Canada; greg.fahlman@NRC-CNRC.gc.ca, peter.stetson@NRC-CNRC.gc.ca.}
\altaffiltext{4}{Centre for Astrophysics and Supercomputing, Swinburne University, P.O. Box 218, Hawthorn, VIC 3122, Australia; bgibson@astro.swin.edu.au.}
\altaffiltext{5}{Division of Astronomy, University of California at Los Angeles, 405 Hilgard Avenue, Los Angeles, CA 90095; hansen@astro.ucla.edu, rmr@astro.ucla.edu.}
\altaffiltext{6}{Observatoire de Strasbourg, 11 rue de l'Universite, F-67000 Strasbourg, France; ibata@newb6.u-strasbg.fr.}
\altaffiltext{7}{Department of Astrophysics, Division of Physical Sciences, American Museum of Natural History, Central Park West at 79th Street, New York, NY 10024-5192; mshara@amnh.org.}

\begin{abstract}
Time-series data taken with the \textsl{Hubble Space Telescope} of a field six core radii ($\sim 5\arcmin$) from the center of the globular cluster Messier 4, and covering a period of about 10 weeks in early 2001, have been analyzed in search of variable objects. Various criteria were employed to select candidate variable stars. Period searches were performed on the selected candidates using phase dispersion minimization (PDM). The reliability of the PDM search results was tested using synthetic light curves of eclipsing binary stars and sinusoidal light curves of different periods. Results from this analysis showed that there are probably no eclipsing binary stars or periodic variables in our field with periods on the order of a few hours to a few days, down to limiting magnitudes of $V\sim 25$ and $I\sim 24$, which is consistent with the absence of contact binaries such as W Ursae Majoris systems. However, one candidate variable star does show an increase in brightness of $\sim 0.1$ magnitudes in both bandpasses, which seems to last for a few days. Possible explanations concerning the nature of this object include a binary system with a white dwarf primary and a low-mass main sequence secondary, or a BY Draconis variable star.  We are able to set an upper limit to the observed fraction of photometric variability in this dataset of $0.05\%$.

\end{abstract}


\section{Introduction}
\label{sec:intro}

It is well known that binary stars are important for the determination of stellar masses.  In addition, knowledge of the population of these objects in globular clusters (GCs) is fundamental to understanding the evolution and dynamical history of GCs.  Even a small fraction of binary stars in a cluster can contribute greatly in preventing or delaying core collapse of the cluster.  A full treatment of this topic can be found in \citet{hut92}.  

This paper presents the results of a search for variability in faint objects in a time-series dataset taken with the \textsl{Hubble Space Telescope} (\textsl{HST}) in a field of the globular cluster M4 (NGC 6121; $\alpha = 16^{h}23^{m}54\fs61$, $\delta = -26\degr32\arcmin23\textit{\farcs}93$ (J2000)).   Even though the the data were  not primarily collected to search for variable stars, the potential for the discovery of such objects did exist, since it included 246 separate observations of the same field over about two and a half months.  These observations represent the deepest ever imaging of a globular cluster, reaching the faintest part of the white dwarf sequence ever observed.  In the context of this project, it allowed for the study of variability within magnitude ranges that have seldom been explored.   Another recent study was performed with \textsl{HST} by \citet{alb01} in which nearly continuous observations for 8.3 days were taken in the core of the globular cluster 47 Tucanae for the primary purpose of searching for planetary transits.  Although no such transits were found, the study did identify 114 variables, most of which are believed to be binary systems.

In the color-magnitude diagrams of globular clusters, observations of a binary star sequence 0.75 magnitudes brighter than the main sequence are rare.  Even in cases where one is found,  many stars in the sequence are actually optical doubles, due to the effects of crowding in globular cluster images.  However, binary stars have been found in globular clusters (see, e.g., \citet{hod92}, \citet{yan94}, \citet{rub96}, and \citet{kal97}).  In fact, W UMa type eclipsing binaries are among the most common type of short-period variable stars found in GCs \citep{ruc00}.  

In addition to binary systems, the data were probed for other types of variable objects, such as pulsating stars, planetary transits, as well as nonperiodic objects such as flare stars, and supernovae in background galaxies.  The study of these objects is vital to research in diverse areas such as stellar populations, cosmology, and the evolution of stars, clusters, and galaxies.

The following section deals with the observations of M4 and the reduction of the data.   Section \ref{sec:expect} discusses what one may expect to find in these data based on previous studies of M4, as well as other globular clusters observed with \textsl{HST}.  Section \ref{sec:candsel} describes the process of selecting candidate variable stars, and section \ref{sec:PDM} discusses how these candidates were searched for periodicity.  Section \ref{sec:addstar} explains how the reliability of results obtained from the period searches were tested.  Section \ref{sec:nonperiod} describes searches for nonperiodic variable stars, planetary transits, and supernovae.  Finally, section \ref{sec:results} presents and discusses results of the analysis.

\section{Observations and Data Reduction}

The data are taken from \textsl{Hubble Space Telescope} (\textsl{HST}) observations of M4 in the program GO 8679 (cycle 9), using the Wide Field Planetary Camera 2 (WFPC2).  Observations were taken over 123 orbits, covering a period of 68.2 days: 1 January - 9 April, 2001, in the natural \textit{HST} photometric filters F606W and F814W \citep{hol95}, which will be simply referred to here as $V$ and $I$, respectively.  The dataset consists of $98\times 1300$ s exposures in the F606W filter ($V$ band) and $148\times 1300$ s exposures in the F814W filter ($I$ band), all taken in the same field of M4, about 6 core radii ($\sim 5\arcmin$) from the cluster center.

\subsection{Preprocessing and Combined-Image Photometry}

The images were preprocessed according to recipes given in \citet{ste87} and \citet{ste98}.  Crowding was not an issue on these images, due to their distance from the cluster center, as well as the high resolution of the \textsl{HST} images.  A master star list, frame-to-frame coordinate transformations, and photometry of the combined image of the field were completed as discussed in \citet{ric02}, which resulted in color-magnitude diagrams that reach apparent magnitudes of $V \sim 30$ and $I \sim 28$.  

\subsection{Single-Image Photometry}
\label{sec:photo}

To create time-series data for each star found in the M4 field, it was necessary to perform photometry on each individual exposure.  To accomplish this, the following steps were taken: (1) the known frame-to-frame transformations were used to create a combined median image of the field for each chip on WFPC2.  This was performed using the MONTAGE2 software written by Peter \citet{ste94}, which transforms the coordinates of each individual image to match a reference coordinate system, and then produces a combined image. (2) Once the combined median image was created, aperture photometry was performed on all objects using DAOPHOT.  Subsequently, point spread function (PSF) photometry was performed using the ALLSTAR program in order to identify stellar centroid positions.  This was done by employing PSFs built specifically for each chip and filter on WFPC2 by \citet{ste01}.  (3) The output photometry and frame-to-frame transformations were then provided to ALLFRAME, again written by Stetson,  in order to perform PSF photometry on each individual exposure \citep{ste94,tur97}.  Once this single-image photometry was completed, the location of any given star and its accompanying photometry was specified for every single exposure taken in the M4 field.

Since \textsl{HST} is above the Earth's atmosphere, it is not subject to seeing effects experienced by its ground-based counterparts.  However, it is susceptible to slight focus changes due to the telescope moving in and out of the Earth's shadow as it orbits, which causes temperature fluctuations (i.e.~breathing).  As a result, the full width half-maximum (FWHM) of the stellar PSFs on the frames can vary slightly between individual images.  The result is that small offsets in the magnitudes calculated from performing PSF photometry will exist from frame to frame.  These must be corrected before performing any analysis on the data.  To accomplish this, the magnitudes that are output by ALLFRAME for every image on a given chip were compared to those of a reference image, which was chosen to be the first image taken in the dataset\footnote{Use of the combined median image as a reference was also explored, but there was a negligible difference found in the final magnitudes and magnitude errors obtained, compared with those found through use of a single exposure as the reference image.}.  The magnitude offset between frames was calculated by taking the weighted mean of the magnitude difference for each star common to both frames.

\begin{figure}
\plotone{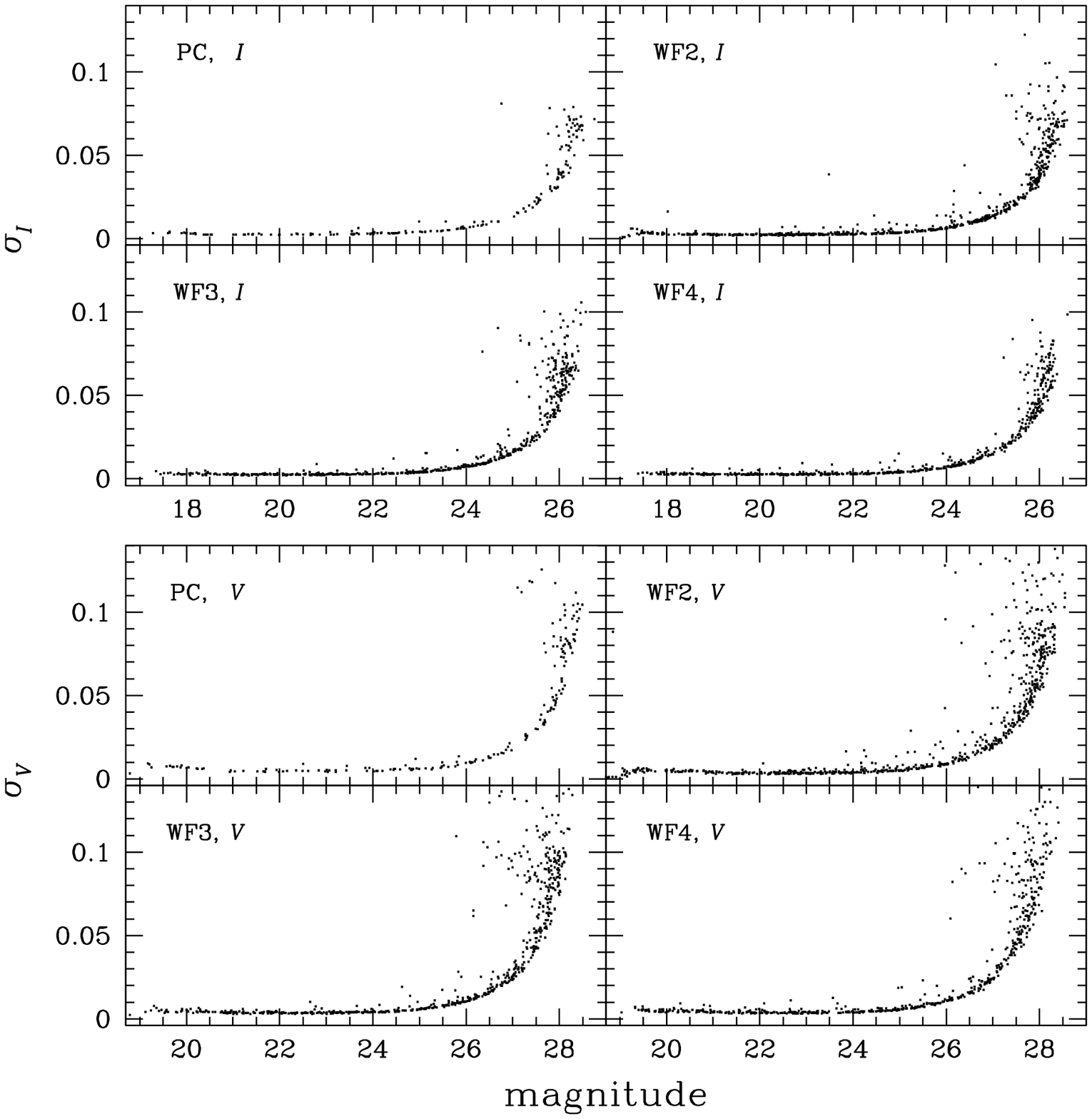}
\caption{Plots of RMS magnitude as a function of magnitude for stars found in each chip of WFPC2 in the M4 images. Each distribution has a well-defined ``elbow'' above which the scatter of data points for the stars rises sharply.  The distribution reaches a maximum in each case at an RMS of approximately 0.1 magnitudes.  The chip and filter are labelled in the top left corner of each plot. \label{fig:rmsall}}
\end{figure}

The usual way of rejecting data points consists of fitting the distribution of magnitudes of a given star with a Gaussian profile, and throwing out any data points at a magnitude greater than $3\sigma$ or so from the mean of the distribution, and perhaps reiterating this process a number of times.  When looking for variability, objects searched may include those which vary greatly in magnitude, such as eclipsing binary stars with deep minima, or supernovae.  By performing this so-called ``$k\sigma$ clipping'' type of data point rejection, data points that may be rejected could actually be caused by real effects due to a significantly varying object.  In order to avoid the loss of potentially useful data, we based the rejection of a given data point on its magnitude \emph{error} rather than its magnitude.  A ``bad data point'' is considered to be one which has a magnitude error that is significantly larger than the rest of the data points for the corresponding star's light curve.   To illustrate how ``significantly large'' was defined, Figure~\ref{fig:rmsall} shows plots of root-mean-square magnitude (RMS or $\sigma$) vs.~mean magnitude in $I$ and $V$ for each chip.  It can been seen that the scatter of data points begins to increase significantly at a well defined ``elbow'' in the distribution, fainter than which the RMS increases rapidly.  With a few exceptions, the maximum scatter of the light curve of any star found in the dataset is approximately $0.1$ magnitudes.  To be conservative, this is the value chosen as the cutoff magnitude error (as returned by DAOPHOT for a single measurement), so that any individual data point with an error above this value was rejected, and any data point with an error below this value was retained.  

\section{Expectations}
\label{sec:expect}

In a past search for binary stars in M4, seven W UMa binary stars thought to be cluster members were found by \citet{kal97} in an $8\farcm8 \times 8\farcm8$ field centered on the cluster with apparent magnitudes reaching as faint as $V = 17.7$.  The field observed in the current dataset is well within this area.  However, unlike the above study, the magnitude range that can be probed for variability in the current study ranges from $20 \lesssim V \lesssim 26$, so all variables found in that study are saturated in the current dataset.  However, there is no reason to expect that the existence of binaries is exclusive to bright magnitudes.  In a recent study of the core of the globular cluster 47 Tucanae (47 Tuc), which probed magnitudes as faint as $V = 25$, 11 detached eclipsing binaries and 15 W UMa systems were discovered in a sample of 46422 main sequence stars, resulting in an observed binary frequency of $0.056 \%$  \citep{alb01}.  In this study of M4, 2102 stars have been analyzed.  If no assumptions are made about the dependence of the distribution of binary systems with cluster radius, one would expect that only one or two binary stars should be discovered in the field observed based on the observed frequency in 47 Tuc.  This already small number is decreased if it is assumed that the majority of binaries are expected to reside close to the cluster center due to mass segregation, since the M4 field is located at about 6 core radii from the cluster center. The picture becomes even less optimistic considering that only 9 of the 15 W UMa eclipsing binaries found in the 47 Tuc study are below the main sequence turn-off point.  However, 71 BY Draconis variables, which are thought to be members of binary systems, were also discovered, making up $0.15 \%$ of the sample.  This converts to a slightly more optimistic expectation of 3 or 4 of these objects being discovered in the M4 field.

As a way of gauging the number of supernova events that are expected to be seen in the current dataset, a comparison is made to a recent study of the Hubble Deep Field (HDF), in which two supernovae were discovered \citep{gil99}.  As with M4, the HDF was observed with the WFPC2 camera on \textsl{HST}, and just as deep, with a limiting magnitude of $I \sim 28$ \citep{fly96,ric02}.  However, unlike the HDF, the M4 images contain zodiacal light, in addition to many stars, due to the cluster's proximity to the Galactic plane and ecliptic.  This results in much higher background levels compared to the HDF, significantly decreasing the number of observable faint galaxies in the field.  As will be discussed in section \ref{sec:sne}, 170 galaxies were identified in the M4 images.  By contrast, $\sim 2100$ galaxies were identified in the HDF \citep{wil96}.  Scaling the discovery of 2 supernovae in the HDF to the current dataset, it is estimated that $\sim 0.08$ extragalactic SNe would be expected in this set of images. 

There also exists the possibility of discovering variable objects which have not yet been discussed, such as cataclysmic variables (CVs), flare stars, transiting planetary-sized objects, or perhaps a new species of variable star.  CVs are very unpredictable, making it difficult to quantify how many are expected to be seen in the M4 field.  However, in globular clusters, which have a high frequency of stellar interactions, it is expected that an abundant population of CVs would be found \citep{dis94,gri01,kni02,tow02}, and indeed, CVs have been found in globular clusters (see, e.g., \citet{sha96,coo98,nei02}).  A relatively small sample of stars may limit this population in the current observations, but the possible existence of CVs in this dataset is not discounted.

The main science driver for the above-mentioned study of 47 Tuc was to define the frequency of close-in gas giant planets orbiting main sequence stars \citep{gil00}.  This study reported a null result, even though simulations that were carried out clearly showed excellent sensitivity to planetary transits.  However, the observations were centered on the core of 47 Tuc, which may be too hostile an environment for planets or protoplanetary disks to survive.  The M4 images were taken far from the core of the cluster, where the number density of stars is much smaller, and thus may provide a more suitable environment for planet formation.  On the other hand, most extrasolar planets are found to be orbiting stars with relatively high metallicity.  M4 is a more metal poor globular cluster than 47 Tuc, perhaps indicating that a smaller fraction of planets will be found in the current dataset. If the null result from 47 Tuc is any indication, it is expected that in the present study of M4, which contains only $\sim 4.5 \%$ of the number of stars in the 47 Tuc dataset, no planetary transits will be found.  In addition, planetary transits found by \citet{cha00}, and very recently by \citet{kon03}, show eclipse depths of $\sim 1\%$ for close-in Jupiter-size planets.  For a relatively bright, non-variable star in this dataset, the scatter in its light curve is typically $\sim 0.014$ magnitudes.  This is most likely too large to confirm a $1\%$ eclipse depth should one exist in the light curve. On a more positive note, however, M4 is known to contain a millisecond pulsar with a planetary companion \citep{ly88,back93,ras94,sig95,thor99,sig03}. The presence of a planet in this unusual system may be suggesting that planets are rather common in at least this globular cluster.

A further consideration is that the stars for which planetary transits have been observed are stars of solar-like mass and radius.  However, the magnitude range of the stars that we have examined for variability is $19\lesssim V \lesssim 26$.  This corresponds to a range in stellar masses of $0.6 M_{\sun} \gtrsim M_{\star} \gtrsim 0.1 M_{\sun}$.  These stars have smaller radii than the Sun, and one would expect that for planetary occulations of such stars, the eclipse depth that would be seen in the star's light curve would be significantly deeper than for the two transiting planets found to date.  In fact, once the stellar mass reaches $0.1 M_{\sun}$, close to 100\% eclipse depths could be expected for a system than is close to edge-on, assuming such a system contains a planet with Jupiter's radius.

\section{Candidate Selection}
\label{sec:candsel}

After reducing the data and rejecting bad data points as described in section \ref{sec:photo}, variability searches were conducted on the light curves of the stars in the dataset.  Described in the subsections that follow are various criteria that were employed in order to narrow down the number of candidate variable stars in this dataset.  After satisfying the criteria, any star that contained in its light curve less than six data points was immediately rejected.  This may seem like a rather generous cutoff value, but when searching for possible supernova events, six points could be enough to determine if it is indeed a viable candidate.  Stars were also excluded from the candidate list if:
\begin{itemize}
\item The image of the star is saturated on the CCD;
\item The star is close enough to the edge of the image so that it can partly appear and disappear over time due to the observational dithering pattern;
\item The stellar image is affected by light from a nearby saturated star; or
\item The star is partially or exactly aligned with a diffraction spike of a bright or saturated star.
\end{itemize}
All remaining candidates that were not excluded due to the above criteria are presented and included in tables accompanying each subsection.  In total, 2102 stars were analyzed for variability.

\subsection{Statistical Outliers}

\begin{figure}
\plotone{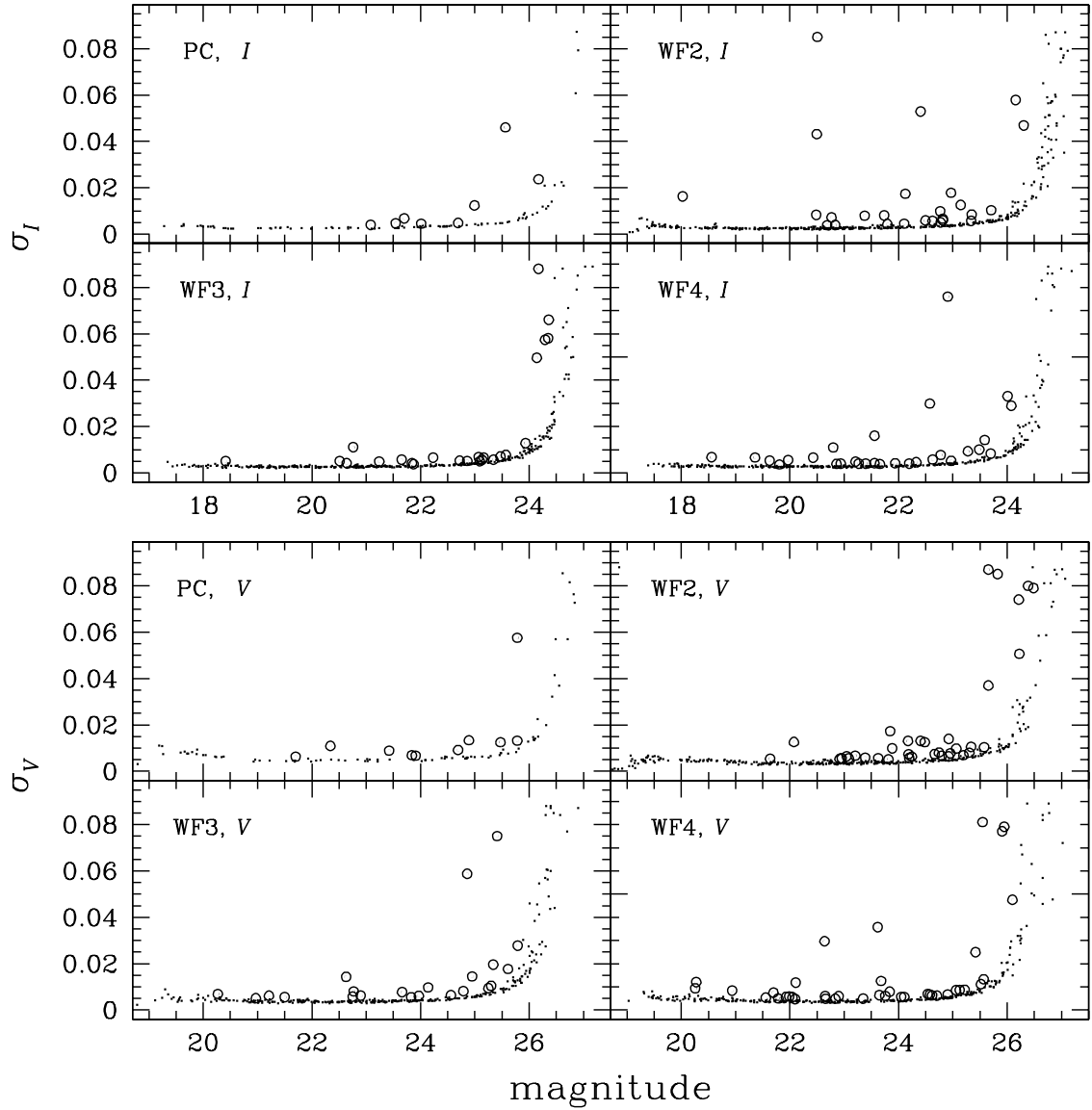}
\caption{Shown as in Figure~\ref{fig:rmsall} (except for rejected stars -- see section \ref{sec:candsel}) are the RMS magnitude distributions for each chip and filter.  Stars chosen as candidate variables are labelled as open circles.\label{fig:rmscandall}}
\end{figure}

A simple identification was first made of outliers in standard deviation (RMS magnitude) as a function of magnitude.  The light curves of these stars have a larger scatter than is expected for stars of similar magnitude, and are thus potential variable star candidates.  These RMS magnitude outlier stars, not including those rejected due to a lack of data as described in section \ref{sec:candsel}, are shown in Figure~\ref{fig:rmscandall}, and candidates chosen by using this method are listed in Table~\ref{tab:rms}.  In this and other tables, coordinates of the objects listed are based on work on M4 by \citet{iba99}, and coordinates for stars not included in that study were calculated using the CCMAP and CCTRAN tasks on the IRAF analysis package.

While the above method of choosing variability candidates based simply on the scatter of the light curves of stars is a good first pass in the candidate search, it is not terribly robust against data corrupted by cosmic rays and warm pixels, which are quite common in \textsl{HST} images.  A more robust candidate selection criterion that attempts to account for these effects is the variability index statistic $J$, a magnitude-independent measure of the confidence that a given star shows real signs of variability \citep{ste96}.  Another attractive feature of this method is that it makes use of data in \emph{all} filters employed in the dataset when determining the variability index.  

\begin{figure}
\plotone{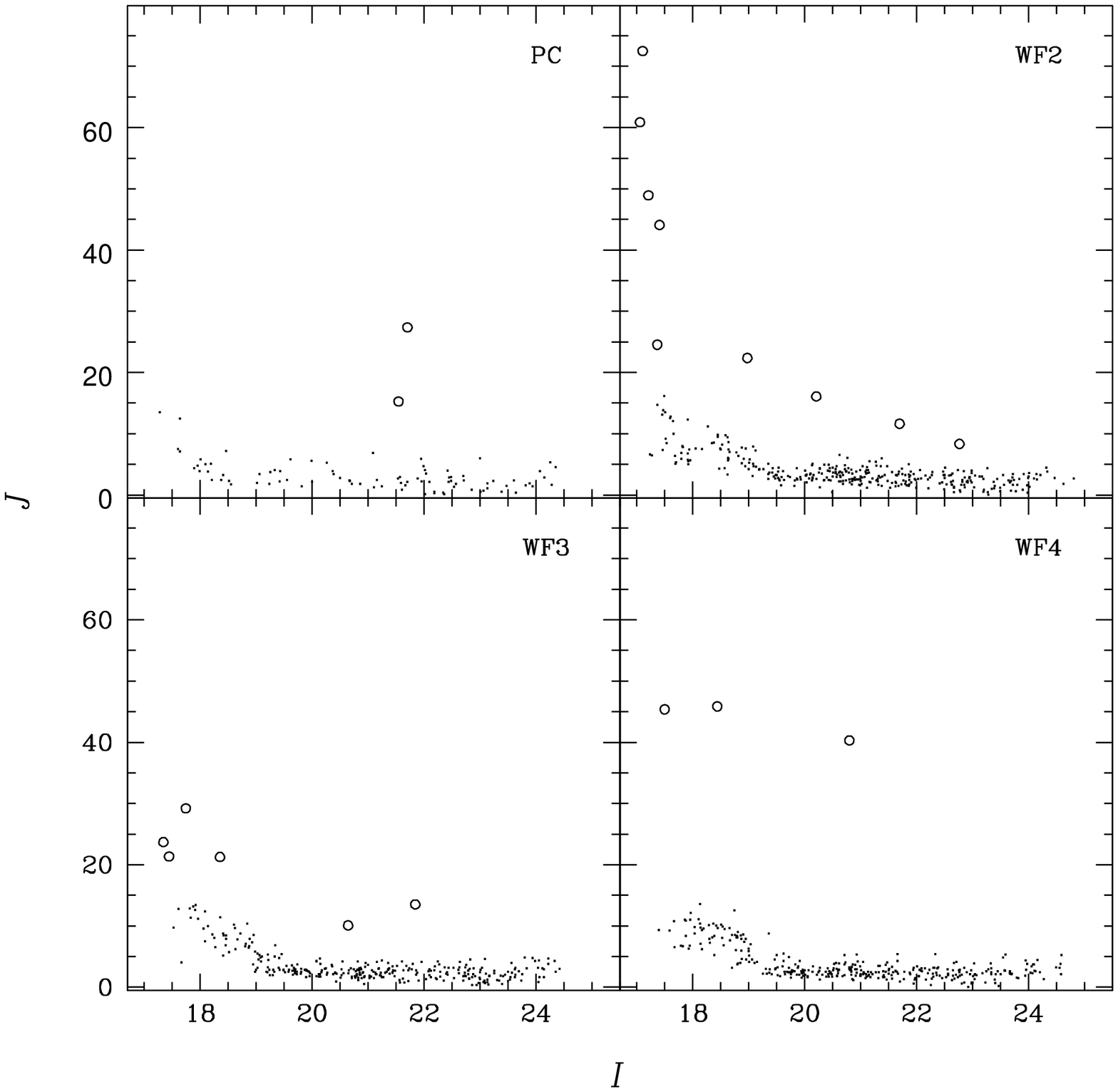} 
\caption{Variability index $J$ as a function of $I$ magnitude for stars found in each chip of WFPC2 in the M4 field.  Outliers to this distribution are plotted as open circles. \label{fig:varcandall}}
\end{figure}

Values of $J$ were computed for every star, and those with values of $J$ that were obvious outliers in the variability index distribution as a function of magnitude were considered variable star candidates.  These distributions are shown in Figure~\ref{fig:varcandall}, where stars selected as candidates are plotted as open circles. None of these stars passed the additional criteria described in section \ref{sec:candsel}, and so no variable star candidates arose as a result of using this method. 

\subsection{Main Sequence and White Dwarf Sequence Outliers}

\begin{figure}
\plotone{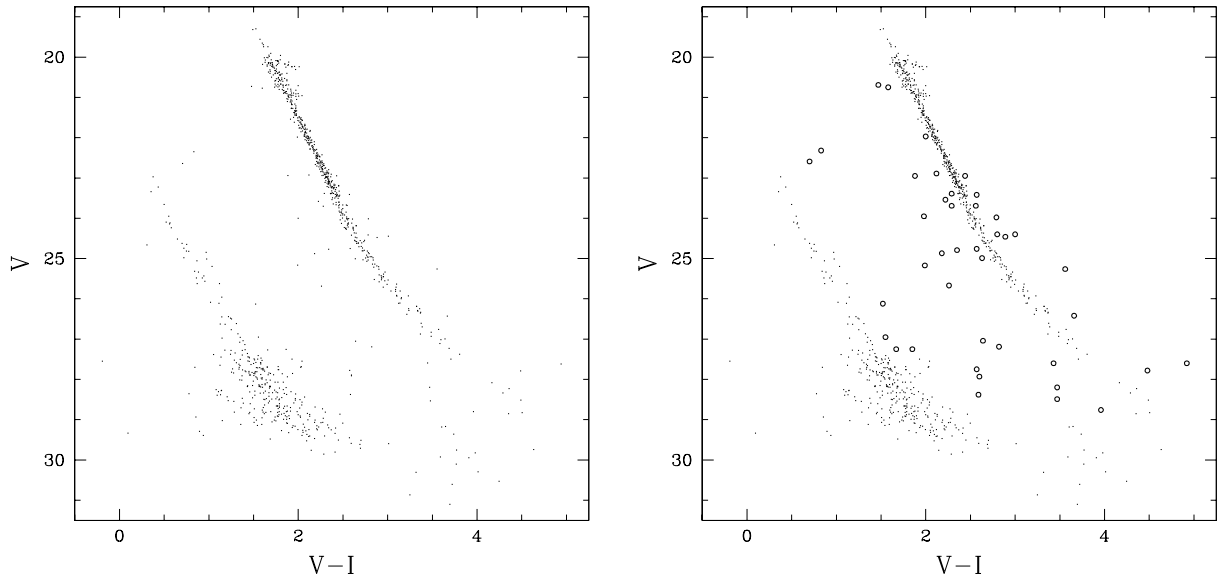} 
\caption{{\it Left}: Proper-motion cleaned color-magnitude diagram of M4.  There are no signs of a binary star sequence running parallel to the main sequence of the cluster.  {\it Right}: Main sequence and white dwarf sequence outliers chosen as variability candidates are plotted as open circles on the same CMD. \label{fig:cmdandoutliers}}
\end{figure}

As explained in Section~\ref{sec:intro}, not many binary star sequences are found parallel to the main sequence in globular cluster CMDs.  Looking at the color-magnitude diagram of M4 (Figure~\ref{fig:cmdandoutliers}, left; \citet{ric02}), one can see no obvious second sequence in the magnitude range observed, other than the expected faint white dwarf (WD) sequence.  Nevertheless, there are stars that appear to be outliers to the main sequence, lying above it, or between it and the white dwarf sequence.  The CMD on the right hand side of Figure~\ref{fig:cmdandoutliers} is the same as that on the left, but stars which obviously do not lie on the MS or WD sequence are plotted as open circles.  From these stars, period searches were performed on those which were not excluded according to the criteria listed in section \ref{sec:candsel}.  These remaining candidates are listed in Table~\ref{tab:outlier}. 

\begin{figure}
\plotone{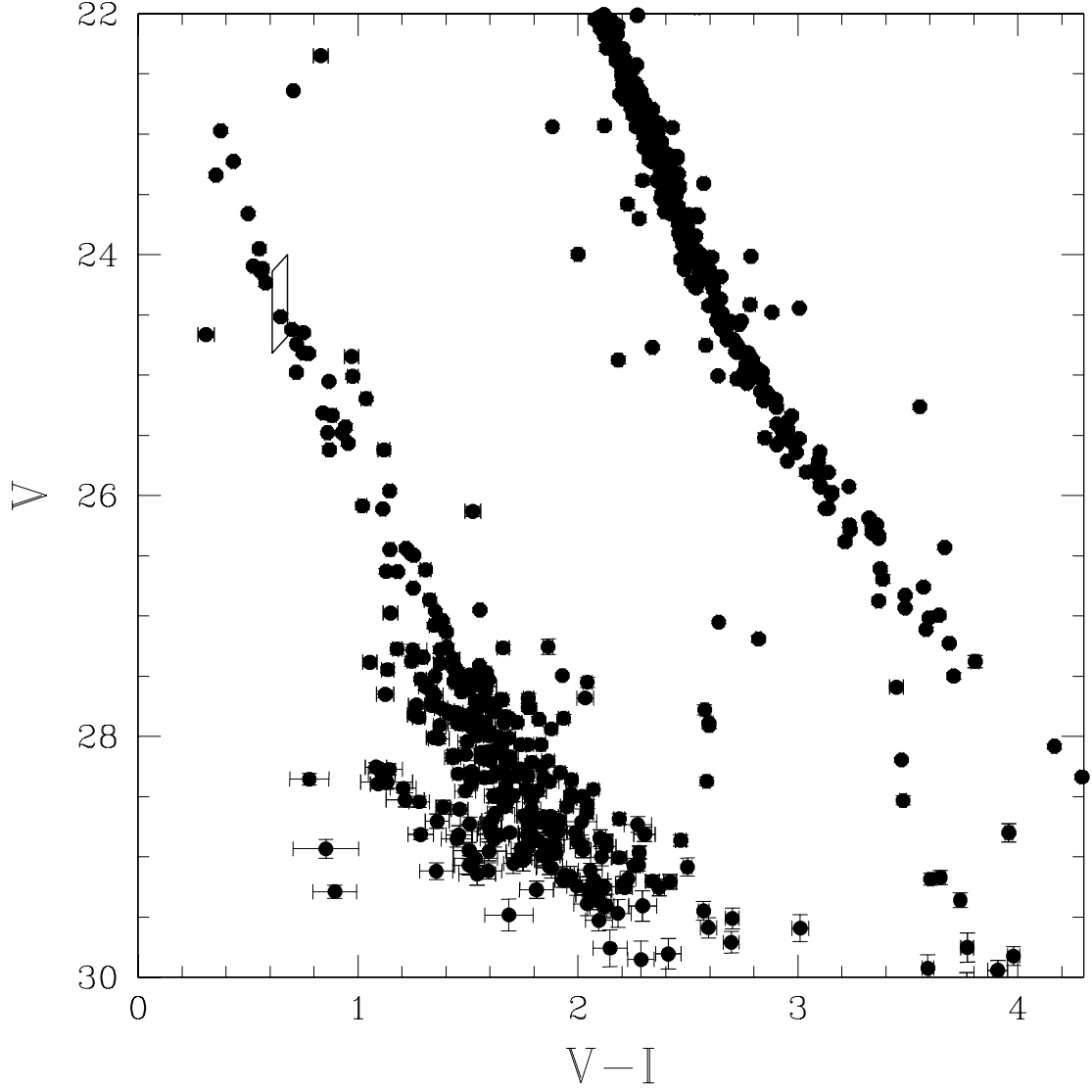} 
\caption{Color-magnitude diagram of M4. The ZZ Ceti instability strip through the white dwarf region is indicated by the box. \label{fig:zzceti}}
\end{figure}

\begin{figure}
\plotone{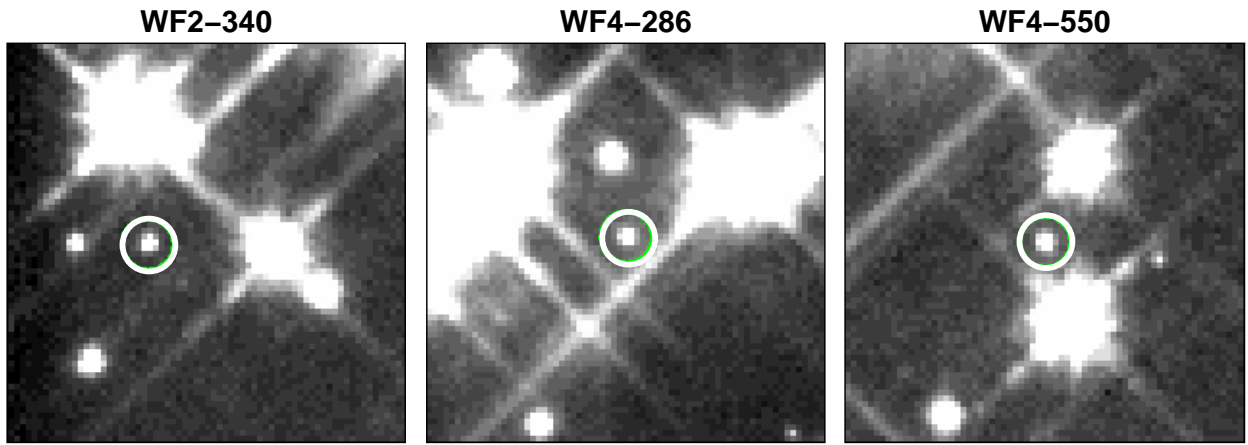} 
\caption{Finder charts for the candidate ZZ Ceti stars in this field. Each chart is $7 \times 7$ arcsec and the potential variables are circled.\label{fig:zzceti_find}}
\end{figure}

The current dataset also includes white dwarf stars that are coincident with the ZZ Ceti instability strip.  Figure \ref{fig:zzceti} shows the M4 CMD, with the ZZ Ceti instability strip bordered by a black box.  Unfortunately, the exposure times of the individual images, which are 1300 seconds for each single observation, are on the order of the typical ZZ Ceti period.  Any variability in these stars would be ``washed out'', since the integration time of each observation is too long to capture the star at various stages in its pulsation cycle.   It is useful however to list the stars which are thought to lie within the ZZ Ceti instability strip, as future observations can be made specifically to study these objects.  Table \ref{tab:zzceti} lists these ZZ Ceti variable star candidates as well as their coordinates, magnitudes and colors, and Figure~\ref{fig:zzceti_find} presents finder charts for these three candidate variables.

\section{Periodicity Analysis}
\label{sec:PDM}

Ever since the landmark paper by \citet{swf78} describing the technique, phase dispersion minimization (PDM) has been used extensively and successfully in the discovery of countless numbers of variable stars and binary systems.  Unlike Fourier transform analysis, which is particularly well-suited for sinusoidal-type variations, it is not sensitive to the shape of a star's light curve.  In addition, its implementation is very straightforward.  However, determination of the statistical significance of results obtained from use of PDM is not so simple, as is the case for Fourier techniques.  It has been shown that the PDM probability distribution does not follow an $F$ distribution \citep{cze97}, as originally thought.  Methods used here to estimate the reliability of findings made by PDM analysis will be discussed in the next section. PDM analysis was performed on all stars in the dataset, and particular attention was paid to candidate variable stars.

\section{Reliability Tests}
\label{sec:addstar}

In an effort to understand the reliability of the PDM analysis performed on candidate variable stars, a number of artificial stars were created by sampling different types of variable star light curves at the same observation times as the M4 dataset.  These artificial stars were then added to the WF2 images, and PDM analysis was performed on the light curves extracted from the resulting images.  

Sinusoidal light curves were simulated with a range of periods, amplitudes, and magnitudes.  The range of magnitudes simulated were $17 \leq I \leq 28$ and $19 \leq V \leq 28$.  The bright end of these ranges correspond to the saturation limits of the dataset in each bandpass.  The faint end corresponds to the faintest magnitude reached in the combined $I$ image.  Six periods were simulated: 0.085, 0.16, 0.42, 2.1, 5.6, and 15.6 days.  The shortest period simulated corresponds to a value just above the Nyquist frequency for this dataset, and the longest period simulated corresponds to approximately the duration of data collection before the occurence of the 44.4-day gap in the observations.  Amplitudes of 0.002, 0.02, and 0.2 magnitudes were simulated.  In all, 20181 sinusoidal light curves were simulated for each filter.

In addition, two types of eclipsing binary star light curves were simulated by employing the widely-used Wilson-Devinney (W-D) code \citep{wd71}.  Using this program, Algol and W UMa-type systems were simulated, each with a 0.42-day period.  The simulated Algol system has an orbital inclination ($i$) of $67^{\circ}$, a relative orbital ellipse semimajor axis length ($a$) of $351.6\,R_{\odot}$, and a primary eclipse depth of $\approx 0.3$ magnitudes.  The simulated W UMa system has $i = 47.5^{\circ}$, $a = 1.88\,R_{\odot}$, and a primary eclipse depth of $\approx 0.2$ magnitudes.  A period of 0.42 days was chosen because it exhibits a case of full phase coverage for this dataset.  In addition, it is within the range of periods typical of close binary stars.  In all, 1922 binary star light curves were simulated and added to the images in each filter.  This light curve data was used in conjuction with the ADDSTAR program in DAOPHOT \citep{ste01} to add artificial stars to the actual M4 images at different magnitudes.  This ensured that the output light curves would have noise levels typical of real stars in the image with similar magnitudes. Photometry was performed on these artificial stars in the same way as on the original images (see section~\ref{sec:photo}).  

\begin{figure}
\plotone{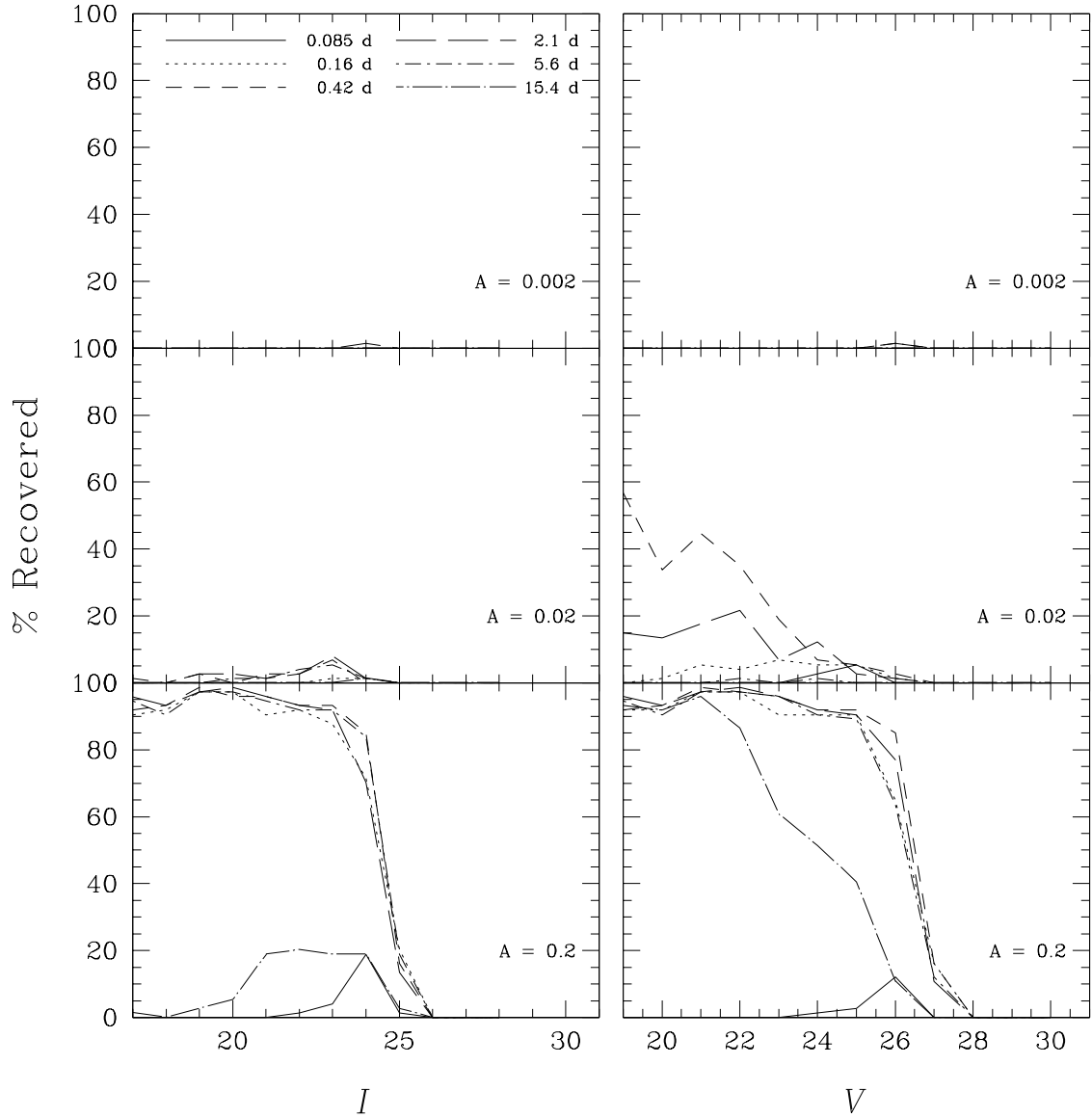}
\caption{Results of PDM analysis on an artificial sinusoidal light curve for a range of periods and amplitudes.  A successful recovery was considered to be within $1\%$ of the input period.\label{fig:simplot}}
\end{figure}

For the artificial sinusoids, a recovered period within 1\% of the correct input period was considered to be a successful recovery.  Results from the simulation of the artificial sinusoidal light curves are shown in Figure~\ref{fig:simplot}.  For the curves in both bandpasses with an amplitude of 0.2 magnitudes, PDM analysis recovers the input period almost $90\%$ of the time down to a magnitude of 24 in $I$ and 26 in $V$ (at which point the recovery rate drops to $\sim 64\%$ at its lowest). There is no dependence of the recovery rate on input period apart from the 0.085 and 15.4-day cases. This gives an idea of the periods to which the PDM analysis is sensitive at this amplitude given the time sampling of this dataset.  For the case of light curves with 0.02 and 0.002-magnitude amplitudes, the ability of PDM analysis to recover the correct input period is significantly lessened.  For an amplitude of 0.02 magnitudes, the recovery rate in $I$-band is very poor regardless of period or magnitude.  For $V$-band, the period search is also discouraging, although the 0.42 and 2.1-day period recovery rate is slightly higher for brighter stars.  In the case of the 0.002-magnitude amplitude curves, the PDM analysis recovery rate is even less successful; the input periods are generally not recovered, regardless of the period or magnitude of the light curve.  This is not surprising, since the scatter of the light curves is on the order of this amplitude or greater (see Figure \ref{fig:rmsall}), even for bright stars.

The results of the tests for the artificial binary star light curves are found in Table~\ref{tab:add_binary}.  These tables show the number of stars added at each magnitude, the number recovered, and the percentage of those stars successfully recovered in each filter.  A recovery was considered to be successful in this case if the period found through PDM analysis is within 1\% of either the correct input period (0.42 days), half the input period, or $\frac{3}{2}$ the input period.  It was found that these are the three major groups within which the periods of the artificial binary stars were recovered.  This is because the light curve folded at half or three halves the input period will cause the primary and secondary eclipses to coincide.  This caused the PDM analysis to occasionally choose these fractions of the input period.  For these artificial binary star reliability tests, the PDM analysis is highly successful in recovering the input periods of the Algol-type light curves down to limiting magnitudes of $I \sim 23$ and $V \sim 24$.  For light curves fainter than this, less than half are successfully recovered.  For the W UMa-type light curves, these limits are $I \sim 24$ ($\sim 65 \%$ recovered) and $V \sim 25$ ($\sim 89 \%$ recovered).   Below these limiting magnitudes, almost none of the stars were recovered with light curves consisting of more than 6 data points (see section \ref{sec:photo}), let alone with the correct input period.

\section{Searches for Non-Periodic Variable Stars}
\label{sec:nonperiod}

In addition to periodic variables,  there may exist objects in the M4 field that vary in magnitude, but not periodically, such as cataclysmic variables, supernovae, and flare stars. There is also the possibility of the presence of objects whose period is too long compared with the time frame in which the observations were taken (or too short relative to the time sampling or exposure times), so that their periodicity may not be evident, as in the case of transits of stars by planets.  Clearly, tools such as PDM are not useful in searches for these objects.  The variability candidates chosen by some of the methods outlined thus far in this section (MS/WD outliers and variablity index, to be specific) \emph{are} valid for these cases and those candidates chosen by these methods were searched for the existence of non-periodic variability in the M4 field.

\subsection{Planetary Transits and Median-Smoothed Light Curves}

As an estimate of the typical period lengths for planetary transits that can be expected to be seen in this dataset, we take an example of a Jupiter-sized planet.  From the reliability tests described in section \ref{sec:addstar}, we can expect to find periodic transits that have eclipse depths of about 20\%.  To get a 20\% dip in the light curve of such a system, the radius of the parent star must be $\lesssim 0.2 R_{\sun}$.  This translates into a mass $\lesssim 0.23 M_{\sun}$ \citep{mon00}, and a magnitude $V\gtrsim 23.2$ for this dataset.  We set a minimum detectable transit time $\sim 5600\mbox{ s}$ based on the minimum separation in time between three consecutive exposures in the light curve.  We estimate the orbital speed implied by such a transit by calculating the length of the transit to be the time it takes for the planet to pass from second contact to third contact.  This implies a maximum orbital speed of $31.6\mbox{ km\,s}^{-1}$.  Assuming a circular orbit ($v_{orb}=2\pi a/P$), the orbital period is then given by:
\begin{equation}
P=\frac{2\pi G(M_{\star}+M_{pl})}{v_{orb}^3} \simeq 71\mbox{ days.}
\end{equation}
This is longer than the length in time over which the dataset was taken, so only one eclipse would be expected to be found.  This is also a lower limit to the period, for the following reasons.  From section \ref{sec:addstar}, the lower magnitude limit at which we successfully recover periods over 90\% of the time is $V\sim 26$.  This corresponds to $\sim 0.1 M_{\sun}$ for main sequence stars.  A lower mass star has a smaller radius, and so the orbital velocity of the planet must be slower in order to have a transit time that can be detected in this dataset, which results in a longer orbital period.  The same argument holds for a higher orbital inclination, since the separation between second and third contacts is smaller.  Thus, periodic signatures of planetary transits would not likely be found for such systems in this dataset.  Searches for potential one-time transits in this dataset are discussed below.

One method used to search for one-time planetary transits and variable star candidates with non-periodic variability involved smoothing the light curve of each star found in the dataset.  This was done by replacing each data point in the light curve with the median of that point and its two nearest neighbouring points.   This has the effect of reducing the scatter in the light curve, so that real variations in the light curve would be more pronounced.  In addition, median smoothing was performed using four and six nearest neighboring points, to search for longer-term variability in fainter stars having noisier light curve data.

This method was executed as follows.  The faintest data point in each median-smoothed light curve was selected, and the ratio of the difference in magnitude between that point and the mean magnitude of the light curve to the RMS scatter of the light curve was calculated.  Outliers to the distribution of this ratio represent data points that deviate more than is expected from the scatter of the smoothed light curve, which would indicate the possibility of a real temporary variation in the star's light curve.  However, all the outliers found were simply due to saturated stellar images, diffraction spikes, or the star being too close to the edge of the image, causing it to disappear and reappear in the frame over time.  Thus, no candidates for planetary transits, or other type of non-periodic variable star, were found using this method.

\subsection{Searching for Supernovae}
\label{sec:sne}

\begin{figure}
\plotone{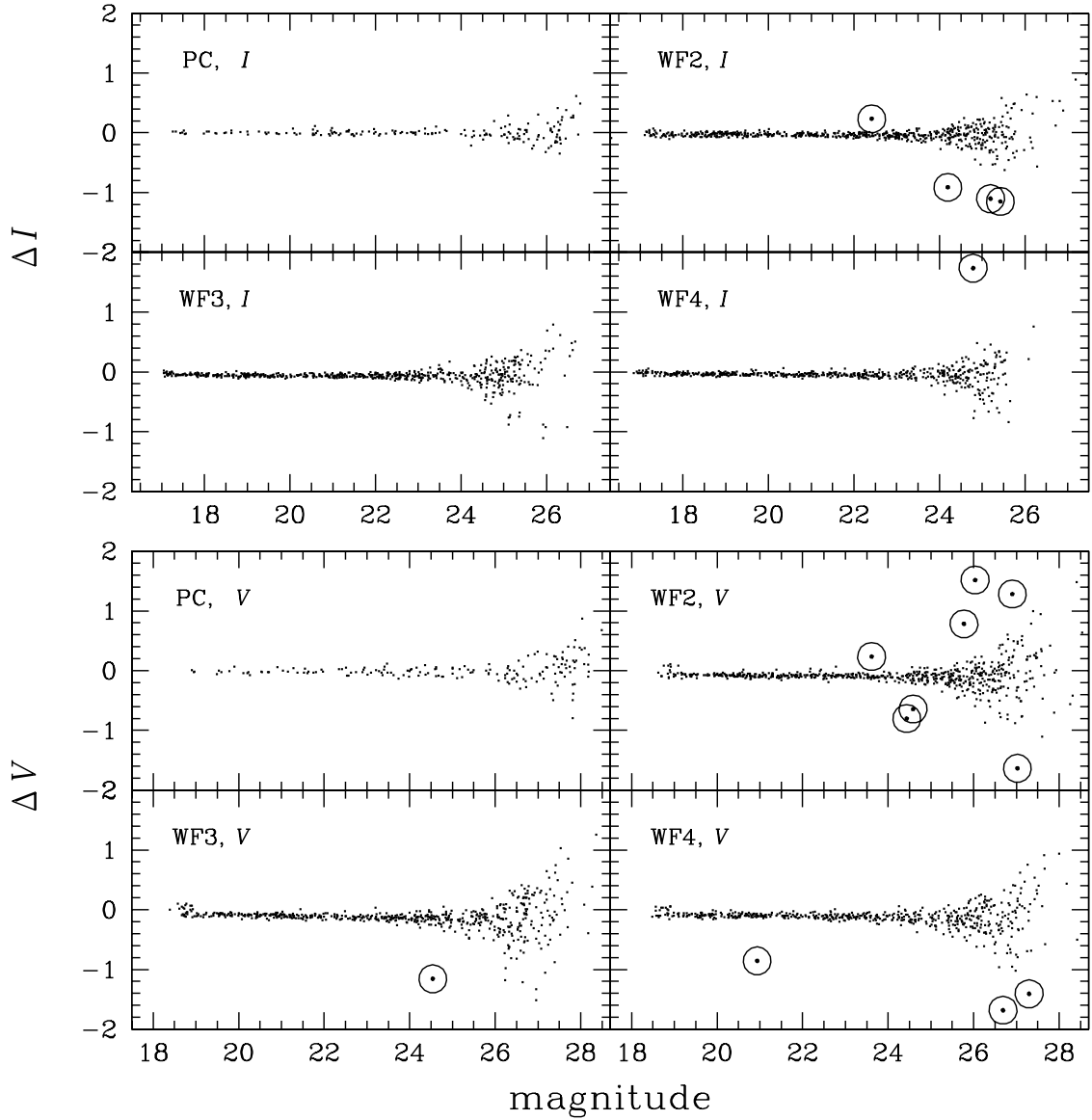} 
\caption{Distribution of magnitude differences between the epoch 1 and epoch 2 combined images.  Outliers to this distribution were chosen as candidate supernovae and are marked by open circles.\label{fig:magdiff}}
\end{figure}

In addition to the median smoothing of light curves, additional techniques were employed to search for possible supernovae (SNe) occuring in the M4 field.  Within the dataset, there is a 44.4-day gap in the data.  For this section, and further discussions pertaining to searching for supernova with this dataset, \emph{epoch 1} will refer to the data taken before this gap, and \emph{epoch 2} to the data taken after the gap.  The data taken during epoch 1 and epoch 2 were combined separately for each bandpass using MONTAGE2 , and PSF photometry was performed on each resulting image using DAOPHOT and ALLSTAR in exactly the same way as described in section \ref{sec:photo}.   The combined epoch 1 images used 100 images in $I$ and 66 in $V$, taken over 14 days. The combined image for epoch 2 used 48 images in $I$ and 32 in $V$, taken over 9 days.   Two methods employing these combined images were used to search for supernova candidates.  The first involved comparing the magnitudes of stars between the two epochs.  Outliers in the distribution of magnitude difference as a function of magnitude, shown in Figure~\ref{fig:magdiff}, were chosen as possible candidates.  However, none of the candidates resulting from this search passed the exclusion criteria described in section \ref{sec:candsel}.  For the second method, background galaxies were identified, and the second epoch image was subtracted from that of the first epoch (after performing sky subtraction and flux normalization for each combined image) in order to search for residual flux at the position of these identified galaxies.  There was no evidence for any residual flux greater than about $1\sigma$ above (or below) the noise levels for the image, except where the nucleus of the galaxy was too bright, in which cases the quality of the subtraction was poor.  However, some of these bright nuclei show stellar-like light profiles, perhaps indicating the presence of active galactic nuclei.  This would make these galaxies interesting objects for possible further study.

\section{Results and Discussion}
\label{sec:results}

After candidates were selected as described in section \ref{sec:candsel}, their light curves were examined more closely and were analyzed using PDM to search for periodicity.   Of all the variable star candidates listed in the tables in the previous sections, only one of these stars (PC, ID \#97) showed real signs of variability in its light curve.  Even for this remaining candidate star, indicators of periodicity were not significant enough to be conclusive.  It is therefore possible that this remaining candidate is non-periodic, or else there is not enough data in the light curve of that star to decisively find a period for it through PDM analysis.  This amounts to an observed photometric variability fraction of $0.05\%$ for this dataset.

\subsection{Remaining Candidate}

\begin{figure}
\plotone{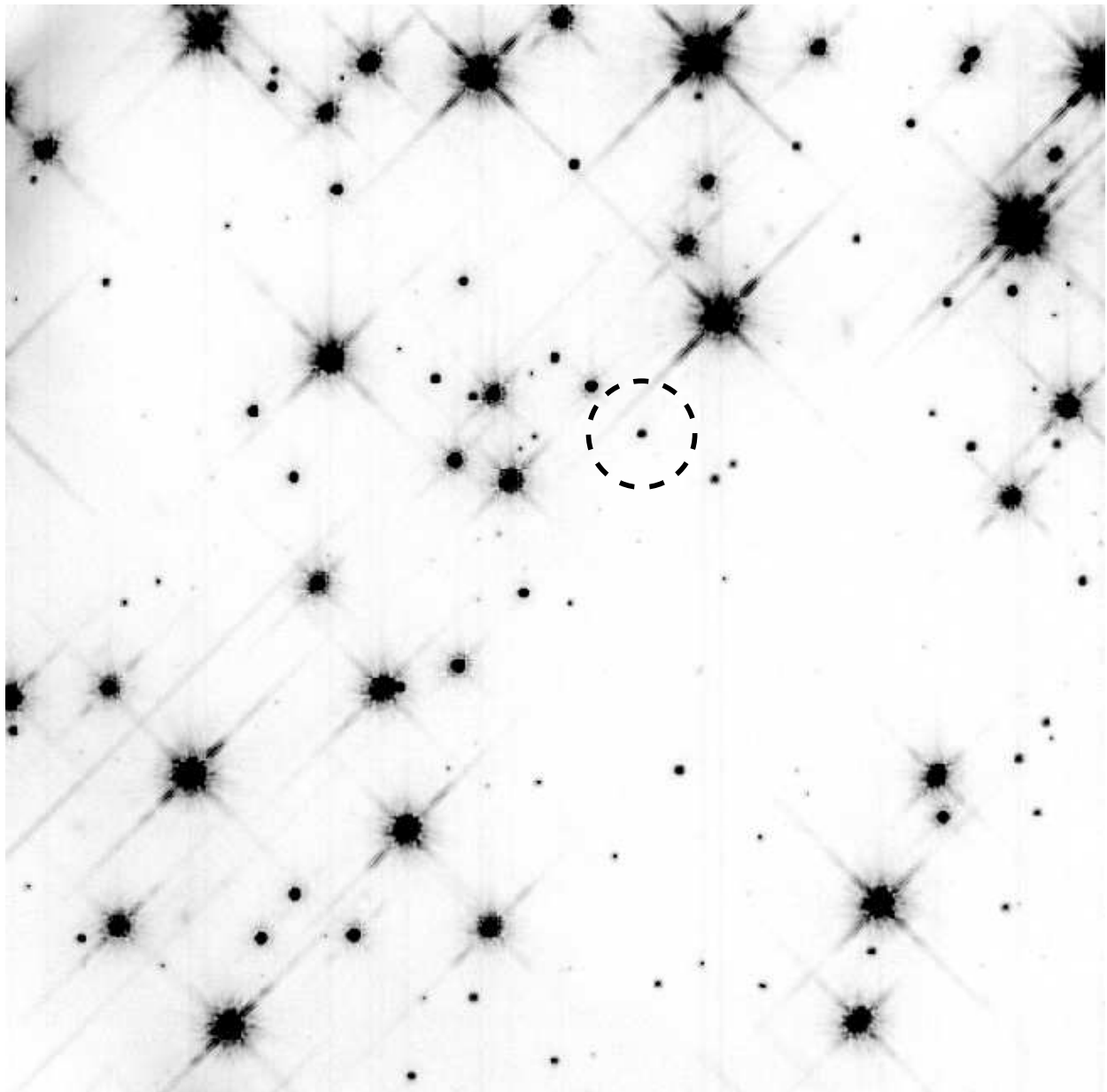}
\caption{The combined $I$-band PC chip image of the M4 field taken with \textsl{HST} ($40 \times 40$ arcsec).   The remaining candidate from the variable star search is enclosed by a dashed circle.  It is a relatively isolated star, in that it is not close to a saturated star or a diffraction spike.  It is also very far from the edge of the image.\label{fig:pc_newI}}
\end{figure}

\begin{figure}
\plotone{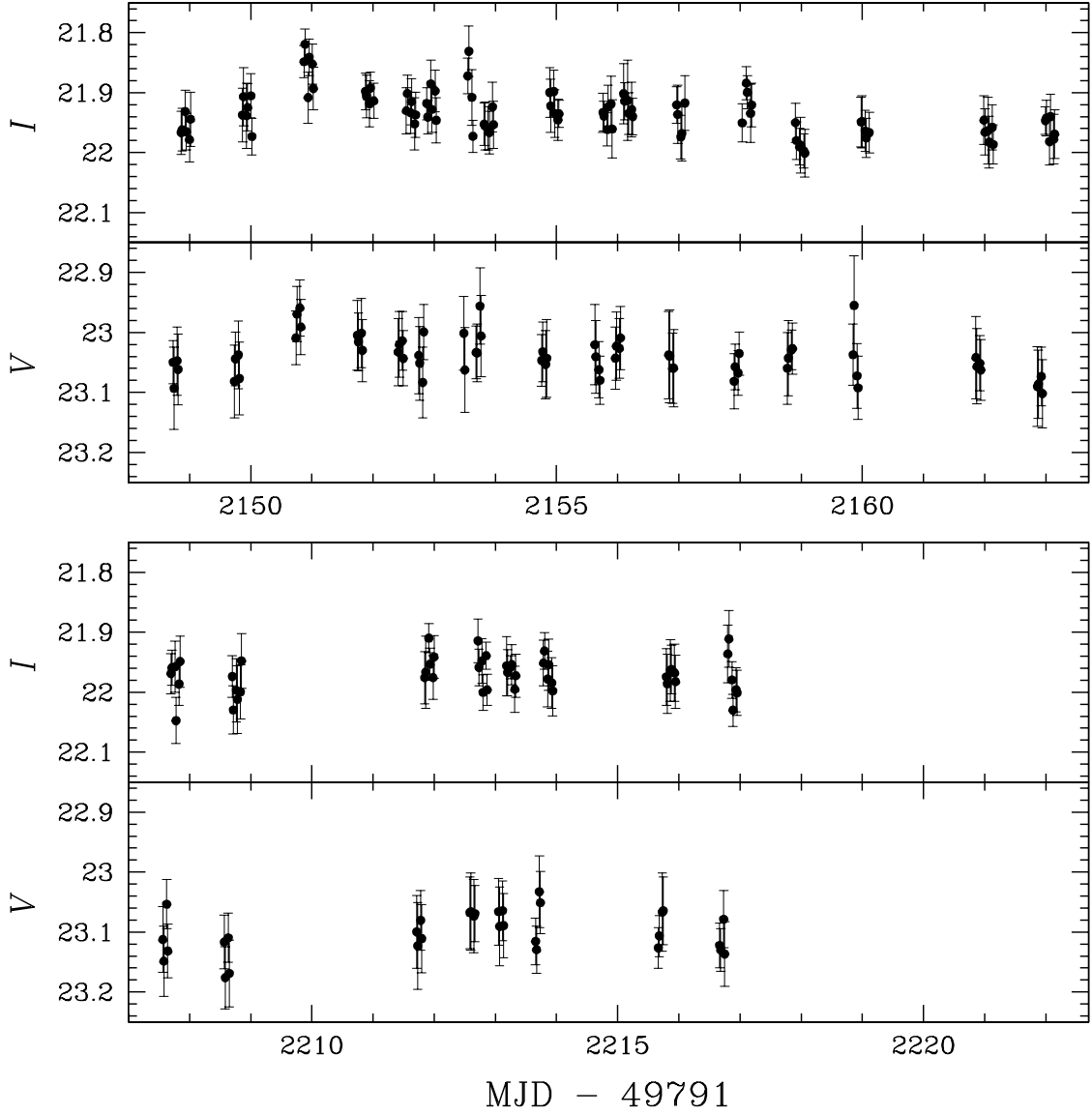}
\caption{Light curve of the remaining candidate variable star.  The first set of rows shows the light curve before the 44.4-day gap in the observations, and the second set shows the light curve after that gap.  It can be seen that the light curve shows a $\sim 0.1$ magnitude rise in its amplitude a few days into the observations in both the $I$ and $V$ filter data, and is a highly desirable characteristic in the selection of this star as a candidate.\label{fig:lastcand_lc}}
\end{figure}

There is the still unresolved issue of the remaining candidate star found in the PC field of the observations.   There is no clear reason that this star should not be considered as variable: it is relatively bright, it is not close to any image defects (such as cosmic rays or diffraction spikes from saturated stars) and is far from the edge of the image.  Figure~\ref{fig:pc_newI} shows a combined $I$-band image of the PC chip observations, in which the star in question is circled.  In addition, it has been identified as a cluster member by its proper motion.  The notable feature in the light curve of this star is the apparent rise in amplitude by $\sim 0.1$ magnitudes within the first few days of the onset of data collection, as seen in Figure~\ref{fig:lastcand_lc}. This peak is seen to occur in both bandpasses at the same time, further supporting the claim that this object is indeed variable.  A clear second peak in brightness is not seen, which could mean that the star is not periodic.  However, it may be that this object has a period such that a second peak occured during a gap in the observations, or else would probably have to be modulated in amplitude over time in order to explain the lack of additional peaks.  Inspection of its light curve seems to indicate a possible period in the range of 5 - 9 days, but periodicity analysis performed on this star showed no significant result.

\begin{figure}
\plotone{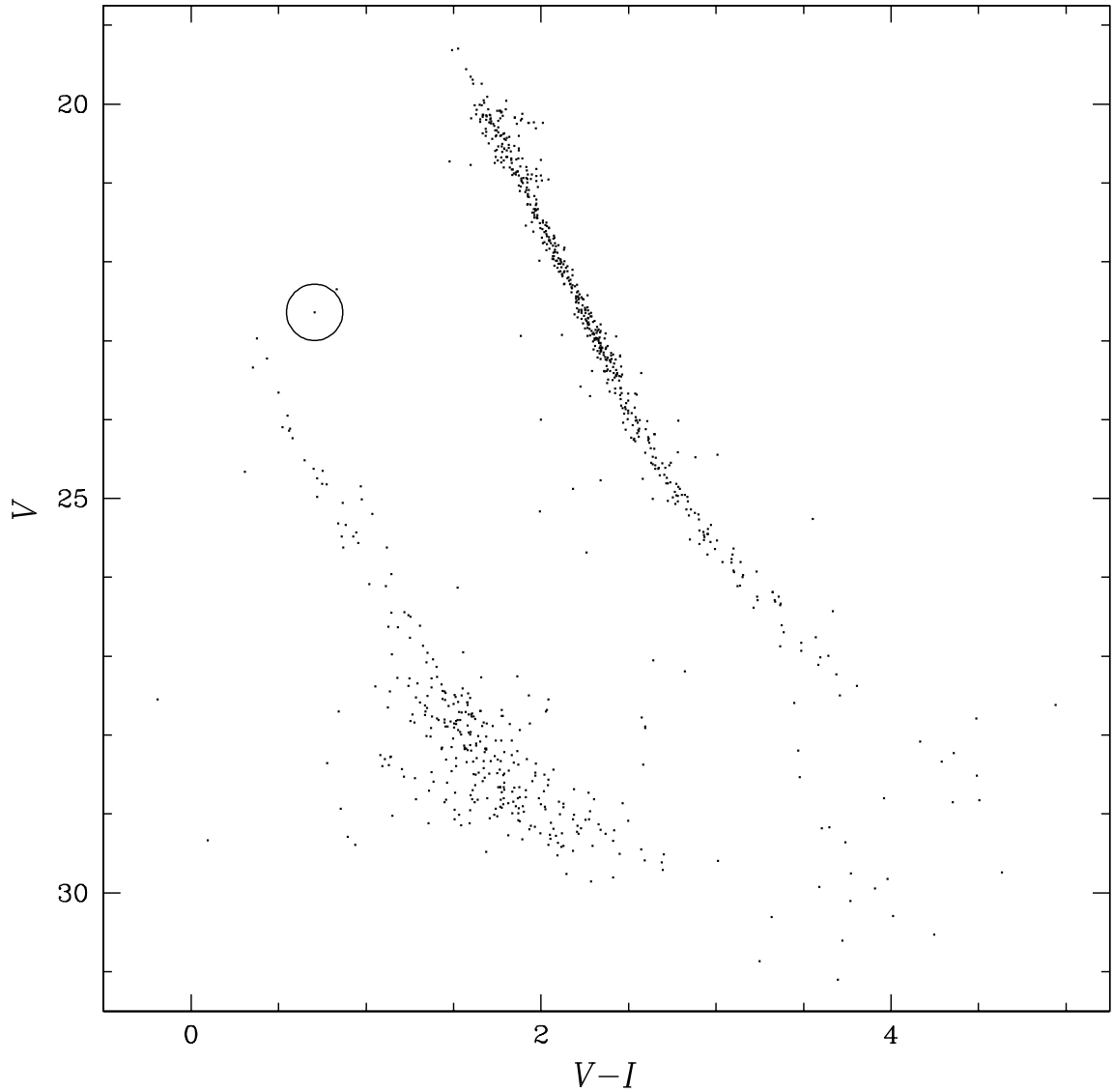}
\caption{The same CMD of M4 shown in Figure~\ref{fig:cmdandoutliers}, with the remaining candidate star circled.  It is at the bright end and slightly to the red of the white dwarf sequence of the cluster.  The slightly more luminous object just to the red of the candidate was very close to the edge of the image, and disappeared from and reappeared in the image over time, and was thus excluded from being chosen as a candidate variable.\label{fig:lastcand_cmd}}
\end{figure}

\begin{figure}
\plotone{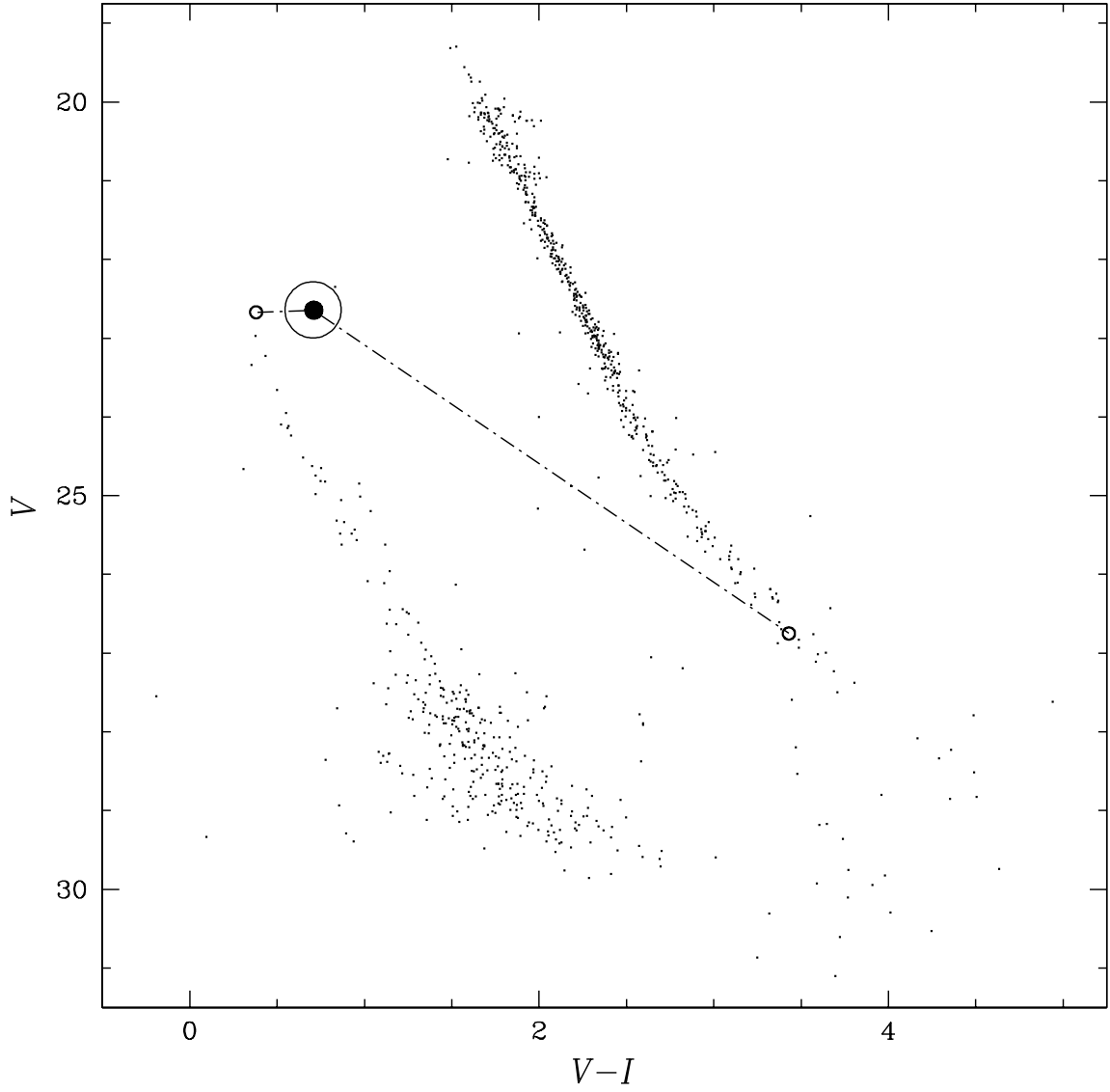}
\caption{Color-magnitude diagram of M4, showing the positions of the two stars (smaller open circles) needed to produce a binary system that would be observed with similar magnitude and color as the candidate star (filled circle).  The resulting object is very close in position to the candidate, which is located at the center of the large open circle.\label{fig:decomp}}
\end{figure}

This star has a magnitude on the combined image of $V = 22.6$ mag and color $V-I = 0.71$ mag, which places it at the bright end and just red of the white dwarf sequence in the CMD of M4, as shown in Figure~\ref{fig:lastcand_cmd}.  The location of this star in the color-magnitude diagram indicates that it may be a binary star, composed of a low-mass main sequence star and a white dwarf.  In order to explore this possibility further, a simple decomposition of this candidate into two stars was performed. We investigated the magnitudes and colors of two separate stars, lying on either the main sequence or white dwarf sequence, that would be needed to produce a single observed star of magnitude and color similar to this candidate.  This gives further insight as to the types of stars that might exist in a binary system found in the position of this candidate on the M4 CMD.  We find that a binary system consisting of a white dwarf with magnitude and color $V \sim  22.7$, $V-I \sim 0.4$ and a faint, late-type main sequence companion star, with upper limits to its magnitude and color of $V \gtrsim 26.8$ and $V-I \gtrsim 3.4$, respectively, result in an object with a position on the CMD that is very near to that of the candidate star.  This is illustrated in Figure~\ref{fig:decomp}.  The positions of the two speculated binary star constituents are shown as the smaller open circles, and the resulting binary is shown as a filled circle.  A large open circle is centered on the position of the actual candidate star.  As one can see, the composite star is very close to the position of the candidate.  Because of the proximity of the candidate variable to the white dwarf cooling sequence, it was expected that such an object would be dominated in luminosity by a white dwarf.  The position of the hypothetical lower main sequence constituent would indicate that this object could be a late M-type star. However, the magnitude and color found for this companion are lower limits, since any companion that is fainter on the main sequence contributes little to the hypothetical binary system.  It is then possible that this companion could be an even fainter, lower-mass main sequence star.  

It may be possible that this object is a cataclysmic variable (CV).  It is thought that such objects can be identified in the color-magnitude diagrams of globular clusters, between the main sequence and the white dwarf sequence \citep{tow02}.  In addition, those that are found close to the white dwarf sequence will be dominated in magnitude by the white dwarf component. The light curve of the object does not seem to show the characteristics of an eclipsing binary system.  However, the possibility does exist that it is a face-on binary system that includes a BY Draconis variable.  The period range of 0.4-10 days of the BY Dra stars found in the 47 Tuc study does not discount this as a possibility.  In addition, that study reported amplitudes for these stars ranging from $V \sim 0.001$ - $0.061$.  This is not much smaller than the $\sim 0.1$ magnitude increase seen in the candidate object's light curve. A further possibility is that the observed increase in brightness is due to flaring on the star's surface, which would not necessarily be periodic, thus explaining the null result of the PDM analysis.  However, this is unlikely, since the timescale typical of stellar flares rising to maximum brightness and diminishing is on the order of hours, whereas the rise in amplitude in the light curve of the candidate star seems to last for days.    In any case, more data would be needed in order to determine the nature of this object's variability.  There is another star next to this candidate on the CMD.  This star (ID \#1), also found on the PC chip image, was very close to the edge of the frame.  Its partial appearance and disappearance from the field over time, due to the dither pattern of the observations, was responsible for the observed variations in its light curve, discounting it as a possible variable star candidate.

\subsection{A Lack of Variability}
\label{sec:no-var}

The lack of eclipsing binary stars in the field can be explained in a number of ways.  Firstly, it may be that any binaries that are present in this field are face-on relative to the line of sight, although this possibility is an unlikely one.  Another reason may be the fact that the observations were taken in a field that is considerably distant from the core of M4.  It is expected that mass segregration in the cluster will, over time, preferentially cause binary systems to sink toward the cluster center, since binary stars are typically more massive on average than single stars. This may not have left many binary systems at the large cluster radius that was observed in this study.  However, a considerable number of $\sim 0.6\, M_{\odot}$ white dwarf stars are seen in the field \citep{han02}, as are main sequence stars up to $\sim 0.65\, M_{\odot}$ \citep{ric02}, so it is not unreasonable to expect, for example, binary star systems comprised of $2 \times 0.3\, M_{\odot}$ stars in that same field.  Another possibility is that due to the high-density environment of globular clusters, low-mass binaries are more prone to disruption before they have the opportunity to harden over time through interaction with third bodies.  Conversely, the relatively large numbers of blue straggler stars in globular clusters (30 have been found in M4 \citep{ruc00}) may represent the coalescence of many low-mass binary systems which have acted as a heat source to the cluster in preventing core collapse.  This could reduce the number of binaries expected in this magnitude range.  However, the real explanation for the lack of eclipsing binaries in this sample may simply be small-number statistics.  As speculated in section \ref{sec:expect}, based on the low fraction of stars in the study of 47 Tuc observed to be in binary systems, the predicted number of these objects recoverable in this dataset was low to begin with.

The small number of background galaxies observable in the M4 images is the most probable reason that no supernovae were found in the field observed.  Previous studies of the Hubble Deep Field, whose images are comparable in depth to the M4 field, yielded the discovery of two supernovae \citep{gil99}.  Due to the substantially larger background flux in the M4 observations compared to that of the HDF (see section \ref{sec:expect}), considerably fewer galaxies were observable, making it unlikely that any supernovae would be found in these observations.

While this research did not produce any confirmed discoveries of variability in the magnitude range $V \sim 19  \,-\, 26$ for M4, it is far from having been fruitless.  Firstly, there is one interesting candidate variable, for which it seems that more observations will be needed to identify its nature.  However, we can place an upper limit to the observed fraction of photometric variability for this study at $0.05\%$, within the magnitude ($19 \lesssim V \lesssim 26\mbox{ and }17 \lesssim I \lesssim 24$), amplitude ($\Delta V \gtrsim 0.1$ magnitude), and period ($0.16\mbox{ days} \lesssim P \lesssim 7.0\mbox{ days}$) ranges to which we were sensitive.

The identification of the white dwarf stars in the M4 field suspected of being ZZ Ceti variables may prove to be of considerable value.  The discovery of such stars in a globular cluster is of the utmost importance to understanding the chemical structure and evolution of population II white dwarfs, which currently suffers from a lack of observational data.

\acknowledgments

We are grateful to Robert E. Wilson for providing the Wilson-Devinney code, and to Puragra Guhathakurta for several very useful discussions on various aspects of this work.  The authors also wish to thank the anonymous referee, who contributed many valuable suggestions.  J.~S.~K.~received financial support during this work through an NSERC PGS-B graduate student research grant.  H.~B.~R.~and G.~G.~F.~are supported in part by the Natural Sciences and Engineering Research Council of Canada. H.~B.~R.~extends his appreciation to the Killam Foundation and the Canada Council for the award of a Canada Council Killam Fellowship. R.~M.~R.~and M.~M.~S.~acknowledge support from proposal GO-8679 and B.~M.~S.~H.~from a Hubble Fellowship HF-01120.01 both of which were provided by NASA through a grant from the Space Telescope Science Institute which is operated by AURA under NASA contract NAS5-26555. B.~K.~G.~acknowledges the support of the Australian Research Council through its Large Research Grant Program A00105171.

\clearpage

\clearpage

\clearpage

\clearpage

\clearpage

\clearpage

\clearpage

\clearpage

\clearpage

\clearpage

\clearpage

\clearpage

\clearpage


\clearpage


\begin{deluxetable}{clllcccccl}
\tabletypesize{\scriptsize}
\tablecaption{RMS outlier candidates for variability search.\label{tab:rms}}
\tablewidth{0pt}
\tablehead{
\colhead{Chip} &  \colhead{ID} & \colhead{RA\tablenotemark{a}} & \colhead{Dec\tablenotemark{a}} & \colhead{Filters\tablenotemark{b}} & \colhead{$V$} & \colhead{$\sigma_V$} & \colhead{$V-I$} & \colhead{$\sigma_{V-I}$} & \colhead{Comments}
}
\startdata
PC   &  1   &  16:23:56.26 &  -26:32:27.0 &   $I$, $V$   &   22.349  & 0.024  &  0.831  & 0.042    &   isolated; very near to  \\ [-3pt]
                                                                                           &&&&&&&&&   edge of image \\
PC   &  2   &  16:23:55.28 &  -26:32:35.8 &   $I$, $V$   &   23.383  & 0.017  &  2.360  & 0.020    &   isolated; close to edge \\ [-3pt]
                                                                                           &&&&&&&&&   of image  \\ 
PC   &  3   &  16:23:54.59 &  -26:32:42.2 &   $I$, $V$   &   24.798  & 0.011  &  2.862  & 0.014    &   isolated; close to edge \\ [-3pt]
                                                                                           &&&&&&&&&   of image  \\ 
PC   &  5   &  16:23:55.56 &  -26:32:32.1 &   $V$        &   23.615  & 0.007  &  1.690  & 0.014    &   isolated                \\ 
PC   &  132 &  16:23:55.10 &  -26:32:04.2 &   $I$, $V$   &   25.969  & 0.033  &  3.295  & 0.036    &   isolated; next to       \\ [-3pt]
                                                                                           &&&&&&&&&   saturated star  \\ 
WF2  &  21  &  16:23:53.41 &  -26:31:29.1 &   $V$        &   24.041  & 0.034  &  2.468  & 0.036    &   next to saturated star; \\ [-3pt]
                                                                                           &&&&&&&&&   very near to edge of \\ [-3pt]
                                                                                           &&&&&&&&&   image \\ 
WF2  &  24  &  16:23:53.56 &  -26:31:29.8 &   $V$        &   24.952  & 0.021  &  2.806  &  0.026   &   next to saturated star  \\ 
WF2  &  73  &  16:23:53.87 &  -26:31:24.1 &   $V$        &   24.040  & 0.021  &  2.552  &  0.027   &   next to saturated star  \\ 
WF2  &  120 &  16:23:54.35 &  -26:31:24.6 &   $I$        &   26.096  & 0.029  &  2.418  &  0.036   &   next to saturated star  \\ 
WF2  &  277 &  16:23:56.44 &  -26:31:31.1 &   $I$, $V$   &   24.097  & 0.036  &  2.839  &  0.046   &   next to saturated star  \\ 
WF2  &  395 &  16:23:59.37 &  -26:31:57.2 &   $V$        &   24.923  & 0.033  &  2.218  &  0.035   &   close to edge of image  \\ 
WF2  &  511 &  16:24:00.26 &  -26:31:48.3 &   $V$        &   25.243  & 0.014  &  3.173  &  0.018   &   near to saturated star; \\ [-3pt]
                                                                                           &&&&&&&&&   close to edge of image \\ 
WF2  &  515 &  16:23:58.25 &  -26:31:10.3 &   $I$        &   25.377  & 0.029  &  3.046  &  0.036   &   isolated; appears to be \\ [-3pt]
                                                                                           &&&&&&&&&   two overlapping PSFs \\ 
WF2  &  585 &  16:24:00.68 &  -26:31:37.1 &   $V$        &   21.197  & 0.024  &  1.368  &  0.031   &   isolated; close to edge \\ [-3pt]
                                                                                           &&&&&&&&&   of image  \\ 
WF2  &  661 &  16:23:56.63 &  -26:32:21.7 &   $V$        &   25.530  & 0.027  &  3.006  &  0.029   &   close to bright star    \\ 
WF3  &  58  &  16:23:59.89 &  -26:33:21.0 &   $V$        &   20.853  & 0.011  &  1.865  &  0.012   &   isolated                \\ 
WF3  &  392 &  16:24:02.42 &  -26:32:59.9 &   $I$, $V$   &   24.117  & 0.036  &  0.565  &  0.039   &   isolated; very near to  \\ [-3pt]
                                                                                           &&&&&&&&&   edge of image \\ 
WF3  &  554 &  16:24:00.53 &  -26:31:49.8 &   $V$        &   21.152  & 0.015  &  1.203  &  0.020   &   isolated                \\ 
WF3  &  611 &  16:24:01.13 &  -26:31:49.2 &   $V$        &   22.406  & 0.025  &  1.349  &  0.026   &   isolated; close to edge \\ [-3pt]
                                                                                           &&&&&&&&&   of image  \\ 
WF3  &  613 &  16:24:01.34 &  -26:31:52.8 &   $V$        &   25.322  & 0.023  &  3.122  &  0.025   &   isolated; close to edge \\ [-3pt]
                                                                                           &&&&&&&&&   of image  \\ 
WF3  &  619 &  16:24:01.02 &  -26:31:45.7 &   $I$        &   19.978  & 0.053  &  1.411  &  0.059   &   isolated; close to edge \\ [-3pt]
                                                                                           &&&&&&&&&   of image  \\ 
WF4  &  11  &  16:23:53.74 &  -26:32:53.0 &   $V$        &   24.614  & 0.033  &  2.711  &  0.036   &   near another star - PSFs\\ [-3pt]
                                                                                           &&&&&&&&&   may be overlapping \\ 
WF4  &  53  &  16:23:56.65 &  -26:32:32.6 &   $I$, $V$   &   24.965  & 0.021  &  2.781  &  0.022   &   isolated; very near to  \\ [-3pt]
                                                                                           &&&&&&&&&   edge of image \\ 
WF4  &  55  &  16:23:56.58 &  -26:32:33.7 &   $V$        &   21.333  & 0.005  &  1.960  &  0.006   &   isolated                \\ 
WF4  &  58  &  16:23:52.32 &  -26:33:13.8 &   $V$        &   21.965  & 0.024  &  2.070  &  0.027   &   next to saturated star  \\ 
WF4  &  63  &  16:23:56.58 &  -26:32:34.8 &   $I$, $V$   &   22.648  & 0.009  &  1.880  &  0.010   &   isolated                \\ 
WF4  &  190 &  16:23:53.79 &  -26:33:16.9 &   $V$        &   24.564  & 0.020  &  2.638  &  0.023   &   next to saturated star  \\ 
WF4  &  345 &  16:23:53.61 &  -26:33:42.4 &   $V$        &   24.964  & 0.022  &  1.436  &  0.023   &   isolated; very near to  \\ [-3pt]
                                                                                           &&&&&&&&&   edge of image \\ 
WF4  &  427 &  16:23:56.32 &  -26:33:32.0 &   $I$, $V$   &   25.599  & 0.016  &  2.735  &  0.020   &   near another star - PSFs\\ [-3pt]
                                                                                           &&&&&&&&&   may be overlapping \\
WF4  &  550 &  16:23:56.72 &  -26:33:48.3 &   $I$, $V$   &   24.235  & 0.030  &  0.580  &  0.032   &   isolated; near to       \\ [-3pt]
                                                                                           &&&&&&&&&   saturated star  \\
WF4  &  551 &  16:23:59.03 &  -26:33:26.9 &   $I$, $V$   &   24.550  & 0.013  &  2.742  &  0.016   &   isolated; near to       \\ [-3pt]
                                                                                           &&&&&&&&&   saturated star  \\ 
WF4  &  555 &  16:23:59.14 &  -26:33:26.1 &   $I$, $V$   &   24.877  & 0.024  &  2.794  &  0.028   &   isolated                \\ 
WF4  &  563 &  16:23:59.19 &  -26:33:26.5 &   $I$, $V$   &   22.291  & 0.031  &  1.431  &  0.035   &   isolated; near to       \\ [-3pt]
                                                                                           &&&&&&&&&   saturated star  \\ 
WF4  &  579 &  16:23:55.60 &  -26:34:02.3 &   $V$        &   22.006  & 0.020  &  2.118  &  0.021   &   isolated                \\ 
WF4  &  581 &  16:23:55.30 &  -26:34:05.6 &   $V$        &   23.336  & 0.022  &  2.413  &  0.029   &   isolated; close to edge \\ [-3pt]
                                                                                           &&&&&&&&&   of image  \\ 
\enddata

\tablenotetext{a}{All coordinates are in J2000.}
\tablenotetext{b}{Filter(s) in which star was identified as an outlier.}

\end{deluxetable}


\clearpage


\begin{deluxetable}{clllccccl}
\tabletypesize{\scriptsize}
\tablecaption{M4 main sequence and white dwarf sequence outliers chosen as candidates for variability search.\label{tab:outlier}}
\tablewidth{0pt}
\tablehead{
\colhead{Chip} & \colhead{ID} & \colhead{RA} & \colhead{Dec} & \colhead{$V$} & \colhead{$\sigma_V$} & \colhead{$V-I$} & \colhead{$\sigma_{V-I}$} & \colhead{Comments}
}
\startdata
PC    &   1     &   16:23:56.26  &  -26:32:27.0  &  22.349 &  0.024  &  0.831 &  0.042  &   isolated; very near to edge of   \\ [-3pt]
                                                                                 &&&&&&&&   image \\
PC    &   97    &   16:23:54.33  &  -26:32:20.6  &  22.639 &  0.009  &  0.706 &  0.015  &   isolated  \\ 
WF2   &   404   &   16:23:56.79  &  -26:31:07.1  &  24.480 &  0.027  &  2.882 &  0.029  &   isolated  \\ 
WF3   &   78    &   16:23:58.95  &  -26:32:58.1  &  25.262 &  0.023  &  3.554 &  0.027  &   isolated; near to saturated star \\ 
WF3   &   532   &   16:24:02.37  &  -26:32:29.9  &  24.014 &  0.023  &  2.786 &  0.027  &   isolated; next to  diffraction   \\ [-3pt]
                                                                                 &&&&&&&&   spike of saturated star \\
WF3   &   625   &   16:24:02.30  &  -26:32:08.4  &  27.190 &  0.027  &  2.821 &  0.029  &   isolated; next to  diffraction   \\ [-3pt]
                                                                                 &&&&&&&&   spike of saturated star; very    \\ [-3pt]
                                                                                 &&&&&&&&   near to edge of image \\
WF4   &   284   &   16:23:54.03  &  -26:33:29.6  &  23.985 &  0.018  &  2.044 &  0.019  &   isolated; near two saturated     \\ [-3pt]
                                                                                 &&&&&&&&   stars \\ 
WF4   &   327   &   16:23:57.94  &  -26:32:59.9  &  23.409 &  0.021  &  2.572 &  0.025  &   isolated \\ 
WF4   &   335   &   16:23:56.22  &  -26:33:16.9  &  22.937 &  0.016  &  1.884 &  0.018  &   isolated \\ 
WF4   &   473   &   16:23:54.87  &  -26:33:53.3  &  24.443 &  0.010  &  3.007 &  0.016  &   next to diffraction spike of     \\ [-3pt]
                                                                                 &&&&&&&&   saturated star \\ 
WF4   &   519   &   16:23:57.81  &  -26:33:31.4  &  22.926 &  0.013  &  2.153 &  0.018  &   isolated; near to diffraction    \\ [-3pt]
                                                                                 &&&&&&&&   spike of saturated star \\ 
\enddata

\end{deluxetable}


\clearpage


\begin{deluxetable}{clllrrcccccl}
\tabletypesize{\scriptsize}
\tablecaption{ZZ Ceti variable star candidates.\label{tab:zzceti}}
\tablewidth{0pt}
\tablehead{
\colhead{Chip} & \colhead{ID} & \colhead{RA} & \colhead{Dec} & \colhead{X} & \colhead{Y} & \colhead{$V$} & \colhead{$\sigma_V$} & \colhead{$V-I$} & \colhead{$\sigma_{V-I}$} 
}
\startdata
WF2  &  340  &  16:23:58.88   &  -26:32:00.0  &  70.567    &   456.524  &  24.515   &  0.025    &   0.649   &   0.028 \\
WF4  &  286  &  16:23:54.13   & -26:33:29.0   &  388.078  &  660.529   & 24.621  &   0.026   &  0.700    &  0.028 \\
WF4  &  550  &  16:23:56.72   & -26:33:48.3   &  749.814  & 482.751   &  24.235   &  0.030    & 0.580    &  0.032 \\
 \enddata

\end{deluxetable}

\clearpage

\begin{deluxetable}{crrcrr}
\tablecolumns{6} 
\tablewidth{0pc} 
\tablecaption{Artificial binary star test results.\tablenotemark{a}\label{tab:add_binary}}
\tablehead{
\colhead{} & \multicolumn{2}{c}{\% Recovered} & \colhead{} & \multicolumn{2}{c}{\% Recovered} \\
\cline{2-3}\cline{5-6}\\
\colhead{$I$} & \colhead{Algol} & \colhead{W UMa} & \colhead{$V$} & \colhead{Algol} & \colhead{W UMa}
}
\startdata
17.0  &  96.0  &  94.6 &  19.0  &  96.0  &  95.9\\
18.0  &  91.9  &  91.9 &  20.0  &  91.9  &  91.9\\
19.0  &  98.6  &  98.7 &  21.0  &  97.3  &  98.7\\
20.0  &  95.9  &  97.3 &  22.0  &  90.7  &  97.2\\
21.0  &  94.6  &  95.9 &  23.0  &  71.7  &  93.3\\
22.0  &  91.9  &  93.2 &  24.0  &  56.9  &  90.5\\
23.0  &  91.9  &  90.6 &  25.0  &  45.8  &  89.2\\
24.0  &  40.5  &  64.8 &  26.0  &  13.4  &  32.5\\
25.0  &   2.8  &   2.7 &  27.0  &   0.0  &   2.7\\
\enddata
\tablenotetext{a}{Primary eclipse depth is $\approx 0.3$ mag for Algol-type stars  and $\approx 0.2$ mag for W UMa-type stars.}
\end{deluxetable}

\end{document}